\documentclass[aps,reprint,superscriptaddress,prb]{revtex4-1}
\usepackage[utf8]{inputenc}
\usepackage{amsmath}
\usepackage{amssymb}
\usepackage{graphicx}
\usepackage{color}
\usepackage[bookmarks=false]{hyperref}
\usepackage[usenames,dvipsnames]{xcolor}
\usepackage{subfigure}
\usepackage[super]{nth}

\begin{document}

\nocite{*}


\title{Superconductor-Insulator Transition and the Crossover to Non Equilibrium in two-dimensional Indium - Indium-Oxide composite}


\author{Bar Hen}
\affiliation{School of Physics and Astronomy, Raymond and Beverly Sackler,\\
Faculty of Exact Sciences, Tel Aviv University, Tel Aviv 6997801, Israel}

\author{Xinyang Zhang}
\affiliation{Geballe Laboratory for Advanced Materials, Stanford University, Stanford, CA 94305}
\affiliation{Department of Applied Physics, Stanford University, Stanford, CA 94305}

\author{Victor Shelukhin}
\affiliation{School of Physics and Astronomy, Raymond and Beverly Sackler,\\
Faculty of Exact Sciences, Tel Aviv University, Tel Aviv 6997801, Israel}

\author{Aharon Kapitulnik}
\affiliation{Geballe Laboratory for Advanced Materials, Stanford University, Stanford, CA 94305}
\affiliation{Department of Applied Physics, Stanford University, Stanford, CA 94305}
\affiliation{Department of Physics, Stanford University, Stanford, CA 94305}

\author{Alexander Palevski}
\affiliation{School of Physics and Astronomy, Raymond and Beverly Sackler,\\
Faculty of Exact Sciences, Tel Aviv University, Tel Aviv 6997801, Israel}


\date{\today}

\begin{abstract}
Magnetic-field tuned superconductor to insulator transition was observed in a novel hybrid system of granular superconducting indium, deposited on indium oxide thin film, which exhibits global superconductivity at low magnetic fields. We have used annealing to tune the coupling to lie just at the borderline where superconductivity in the underlying InOx is suppressed, which is also close to the metal-insulator transition of the InOx. The hybrid system exhibits a "giant'' magnetoresistance above the H-SIT, with critical behavior that manifests the duality between Cooper pairs and vortices.
\end{abstract}


\maketitle

\noindent {\it Introduction} -
Two-dimensional (2D) superconductor to insulator transitions (SITs) tuned by varying disorder, carrier density or magnetic field, have received much attention because they allowed for detailed exploration of the nature of these classes of quantum phase transitions (QPTs) as well as their proximate ground states \cite{Hebard1990,Yazdani1995, Ephron1996,Gantmakher1998,Sambandamurthy2004,Steiner2005,Markovic1998,Bielejec2002,Baturina2004}. In particular the magnetic-field tuned SIT (H-SIT), appears to be broadly relevant to other classes of QPTs such as the quantum-Hall to insulator transition (QHIT) \cite{Sondhi1997}, while also can shed light on the anomalous transport behavior observed in many unconventional superconductors such as the high-Tc cuprates \cite{Steiner2005}.  Here we identify a superconducting-state if the resistivity at low-$T$ is an increasing function of $T$ and insulating-state if a decreasing function. However, in many experiments the superconducting phase ``fails'' to realize, revealing an anomalous intermediate metallic phase. In fact, most amorphous films exhibit such an intermediate metallic phase (for a recent review see: \cite{KKS2018}), even though sometimes its occurrence can be pushed to lower temperatures after improved experimental conditions \cite{Tamir2019}, while some unique material systems, notably indium-oxide \cite{Hebard1990,Sambandamurthy2004,Steiner2005,Breznay2016} and titanium-nitride \cite{Baturina2004} seem to exhibit a ``true'' H-SIT.

A key issue in the exploration of the phases and phase transitions that may occur in 2D disordered superconductors relates to the morphology of the material-systems. For example, the normal-state resistance is not a good predictor for the universality class of observed SIT in different material systems. For example, the very nature of the transition, which involves the integrity of Cooper-pairs, and the creation and ability to pin vortices suggest that different-scale inhomogeneities, which can only weakly affect the normal-state resistivity must be taken into account. This consideration becomes even more severe in the superconducting state, where even weak microscopic disorder, let alone stronger inhomogeneities, are magnified to result in appreciable gap inhomogeneities (see e.g. \cite{Howald2001} for an example in high-Tc cuprates).  It is therefore desired to study a `true' H-SIT in a system with controlled inhomogeneities.  Indeed, in an attempt to elucidate the main features associated with the different scenarios, B{\o}ttcher {\it et al.} \cite{Bottcher2018} have used a highly regular and controllable array of patterned superconducting Al islands connected by a quantum well, where the inter-islands coupling could be gate-tuned to exhibit superconducting, metallic, and insulating behavior. Since this study was limited to an ordered array of 40$\times$100 islands, It is of utmost importance to extend this work to a macroscopic random array of superconducting grains.

In this paper we focus on demonstrating the importance of inhomogeneities by superimposing a granular structure on the otherwise heavily studied uniform amorphous-indium-oxide (InOx) system \cite{Hebard1990,Gantmakher1998,Sambandamurthy2004,Steiner2005}. Here thin film made of continuous amorphous insulating InOx are covered with discontinuous layer of superconducting indium islands. The underlying InOx film is chosen to have resistance that while insulating, can still allow for weak proximity coupling between the almost touching pure indium islands. Thus, while superconductivity originates from the indium grains, it is strongly affected by the InOx film, which we will show is a major player in the resulting SIT, especially in determining the behavior above the transition in the insulating state. Contrasting this new system with previously studied SIT platforms, we are able to point to the key effects of the imposed granularity.\\

\begin{figure}[ht]
	\centering
	\includegraphics[width=1.0\columnwidth]{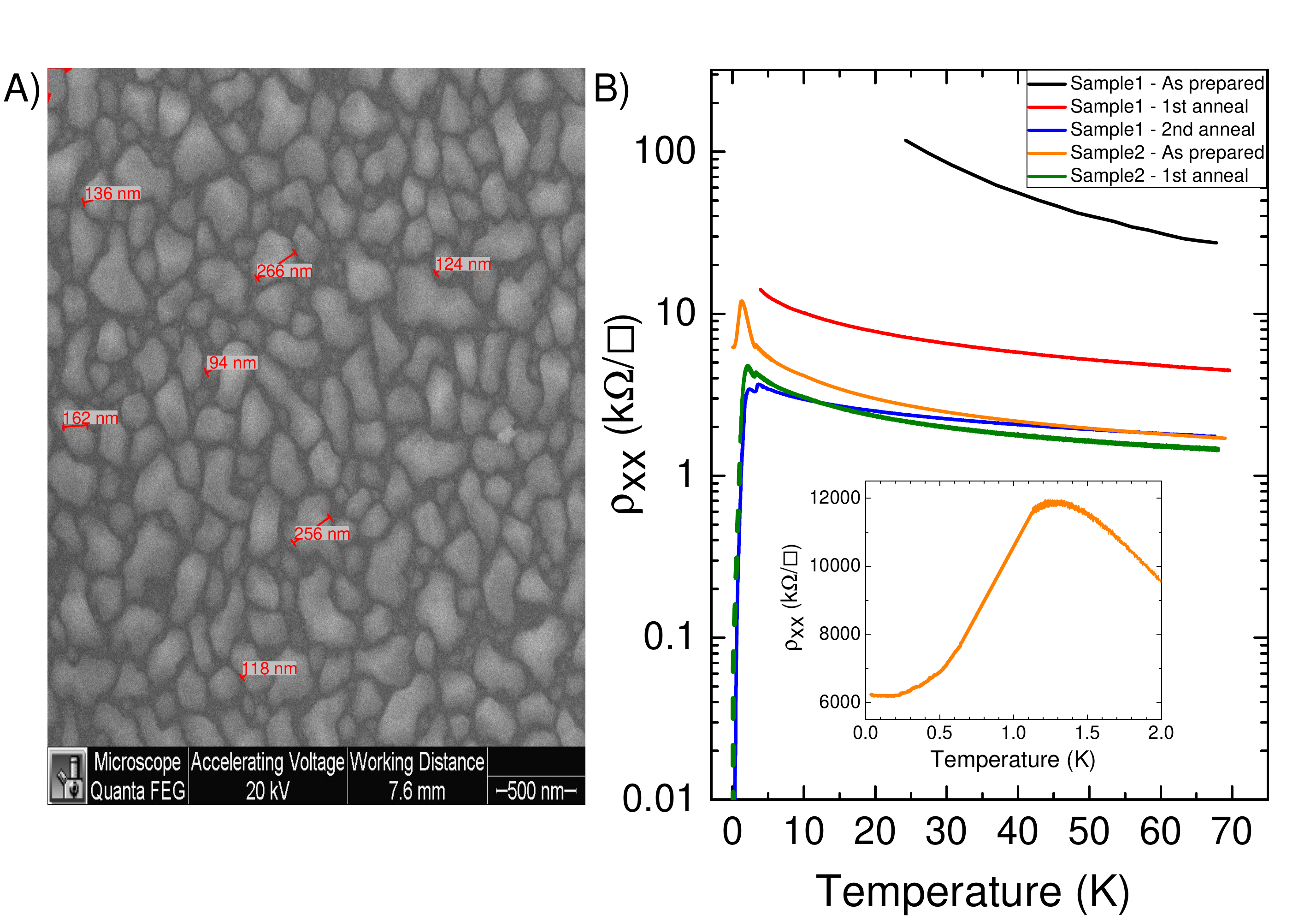}
	\caption{Sample characterization. (A) SEM image of Sample1 confirming the granular (discontinuous) structure of the In on top of the amorphous InOx. The average size of the grains is 1500 Å. (B) Film resistance versus temperature for both samples. The desired range of the resistance is obtained by tuning the inter-grain coupling using annealing. The inset shows the low temperature dependence of Sample2-As prepared. While this sample shows a drop in resistance towards a superconducting state, it fails to complete this transition, ending up saturating at a value close to h/4e2 (this regime of coupling is not discussed in the present paper). The 1st anneal seems to exhibit a superconducting transition very close to T=0. }
		\label{sem}
\end{figure}

\noindent {\it Experiment} -  As mentioned above, our system consists of in-situ evaporated In islands on top of thin films of InOx. For our study we employed two different substrates. Semi-insulating GaAs was used for Sample1 was whereas Lithium-ion glass ceramics (Li-ICGC; from MTI corp.) substrate allowing electrostatic gating was used for Sample2.
A 300 \AA \hspace{0.1cm}continuous layer of amorphous InOx was deposited onto these substrates using an In$_2$O$_3$ pallet, in an oxygen atmosphere at pressure of $5\times10^{-5}$  Torr and at a slow rate of $\sim 0.5-1$ \AA/sec. Then, the layer was covered by In islands by evaporating pure Indium, at base pressure of $5\times10^{-7}$ Torr and at a rate of  $\sim 20$ \AA/sec
The amorphous structure was confirmed by SEM imaging of the film (Fig.~\ref{sem}(A)). The average size of the crystal In islands was found to be 1500 \AA.

Contacts were made using e-beam evaporation of Au-Ti on the corners of the samples enabling transport measurements in Van der Pauw configuration. Aiming to obtain InOx films on the insulating side of the metal-insulator transition (MIT), as prepared films showed an immeasurably high values of resistivity at low temperatures. Heating in vacuum changed the oxygen concentration in the InOx without affecting the In islands, thus allowing for control over the conductance of the intergrain coupling, which in turn determined the overall transport response and in particular the nature of the H-SIT. The annealing process was used for several InOx platforms, however, some of the more dramatic SIT effects were obtained for a combined room-temperature sheet resistance of $\rho_N \sim 0.9 $k$\Omega/\Box$, which is the type of samples we discuss in this paper (Fig.~\ref{sem}(B)). Subsequent transport measurements were performed in a dilution refrigerator in a wide temperature range (28 mK - 70 K) and in magnetic fields up to 1 T which was sufficient to suppress the superconducting state in the samples.

\noindent {\it Results} -  The results presented below are focused on low temperature electronic transport of Sample1-\nth{2} anneal and Sample2-\nth{1} anneal both being the closest to the SIT critical point from the superconducting side.

\begin{figure}[ht]
	\centering
	\includegraphics[width=1.0\columnwidth]{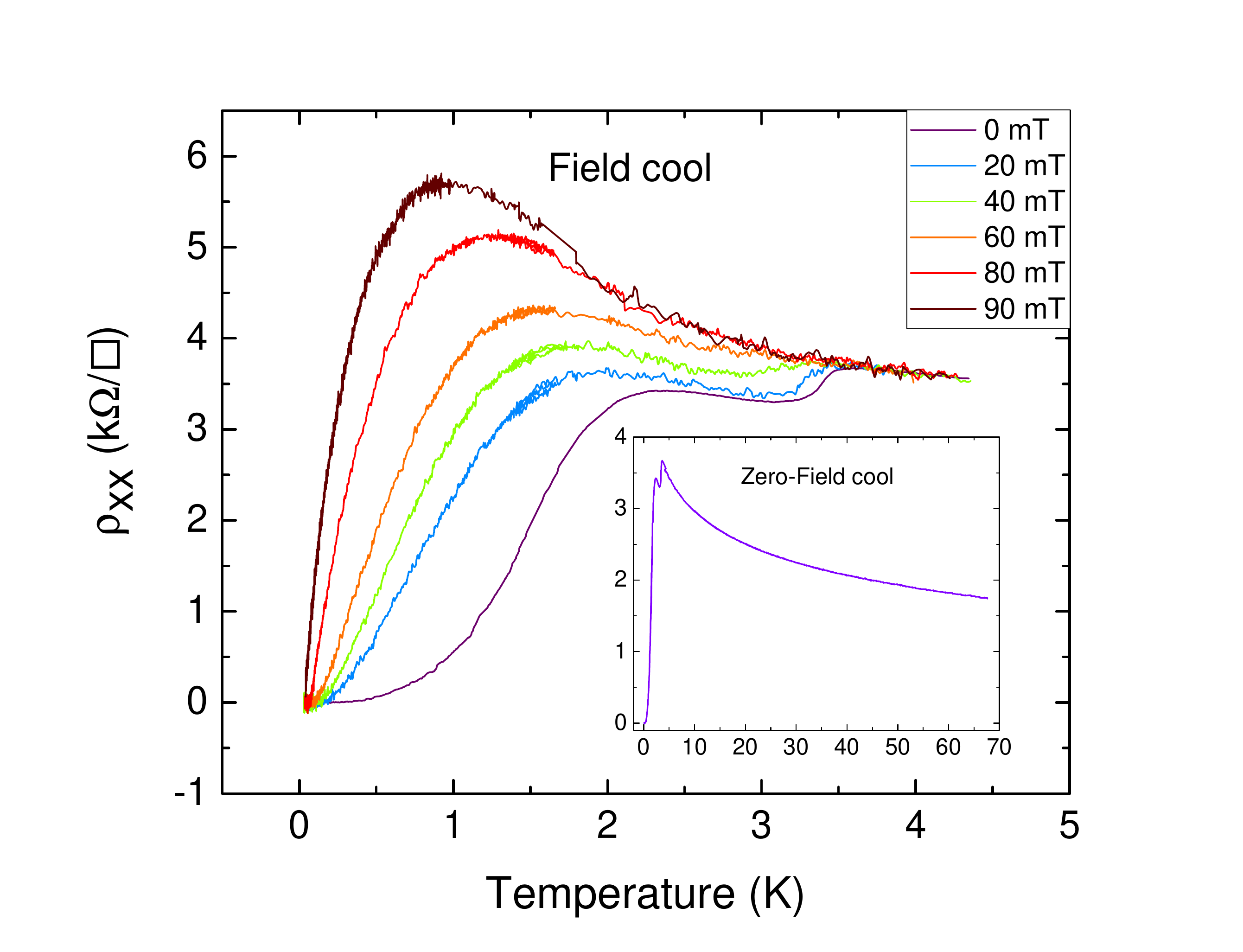}
	\caption{Low-temperature cooling curves at low magnetic fields for Sample1-\nth{2} anneal. Low-temperature cooling curves at low magnetic fields. Two step transition is clearly observed at very low fields. The inset depicts the insulating behavior at zero magnetic field and higher temperatures followed by the superconductor transition.}
		\label{RvsTH}
\end{figure}

The cooling curve (inset of Fig.~\ref{RvsTH}) of Sample1-\nth{2} anneal shows a negative temperature coefficient of the resistivity (TCR, $d\rho/dT<0$) corresponding to an insulating behavior in the temperature range of 70 K$-$3.4 K. Then, at the transition temperature of pure indium, $T_c^{{\rm In}}=3.4$ K the resistivity drops by $\sim10\%$ followed by a slight increase which is attributed to the insulating behavior of the InOx.  For temperatures below $\sim2 $K the TCR changes sign again resulting in a full superconductivity of the film with zero resistance below 0.5 K. This sample is the closest to the SIT from the superconducting side in our set.

Low-temperature cooling curves of Sample1-\nth{2} anneal at low magnetic fields are shown in Fig.~\ref{RvsTH} and further reveal the nature of the double peak transition. While at $H=0$ the first peak clearly coincides with $T_c^{{\rm In}}$, this feature disappears and merges to the lower temperature curve at $\gtrsim 70$ mT. This is more than twice the critical field of pure type-I superconducting In, demonstrating the decrease in mean free path in the thin film form of the material \cite{Chaudhari1966}, particularly reflecting the nanostructured nature of the individual grains, which were demonstrated to reach a critical field as large as $170$ mT \cite{Mikhailin2018}. Thus, the disappearance of the resistance drop upon the increase of the magnetic field may not indicate the actual disappearance of superconductivity in the individual In islands, but rather suggests that the positive TCR of the superconductivity of In is compensated by the negative TCR of the underlying InOx. Further increase of the magnetic field recover the normal-state insulating behavior before a global decrease in the resistance as the combined system seems to tend towards a superconducting state at $T=0$.

\begin{figure}[ht]
	\centering
	\includegraphics[width=1.0\columnwidth]{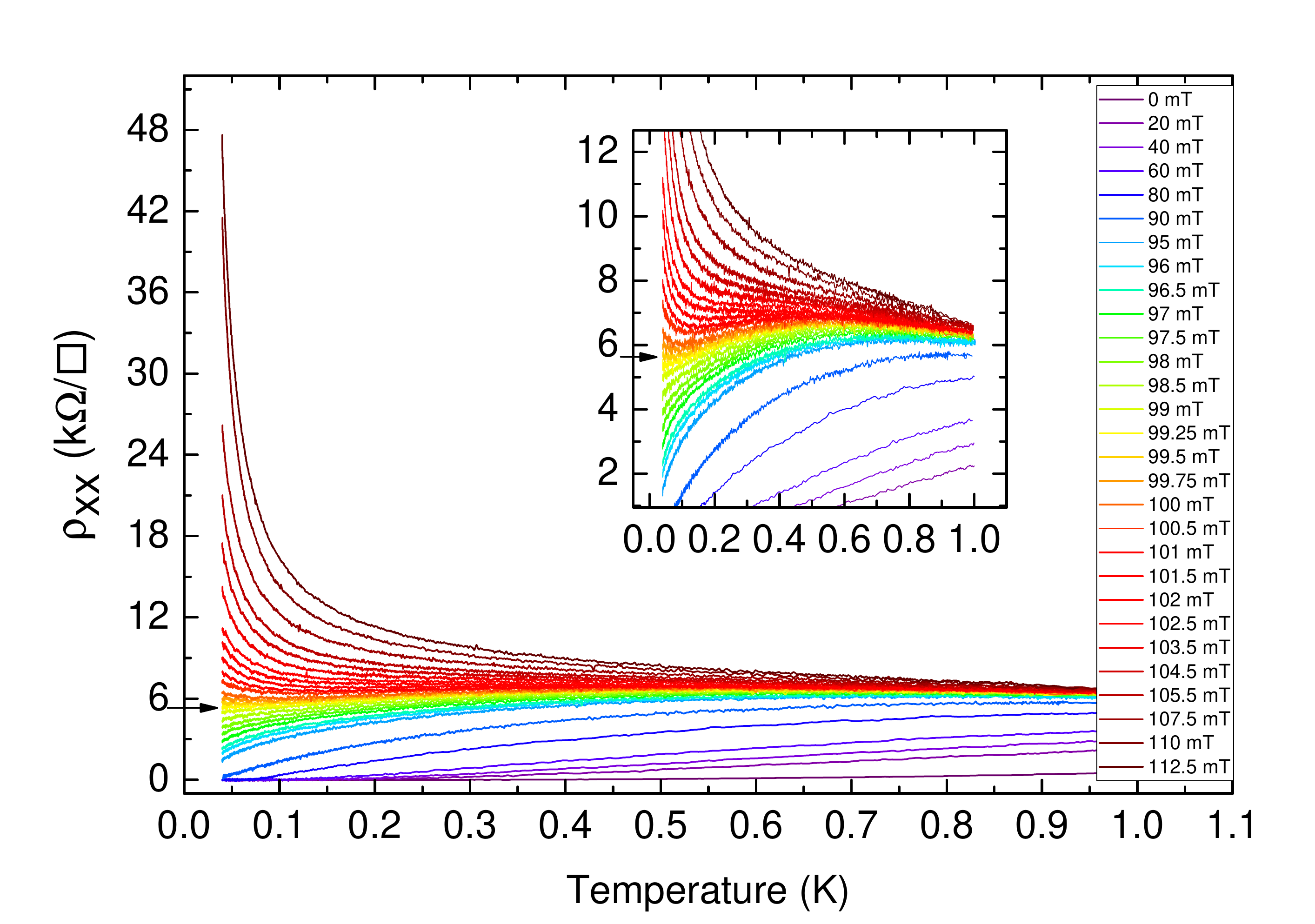}
	\caption{Series of cooling curves of Sample1-\nth{2} anneal at different magnetic fields indicating the SIT. The field which separates the superconducting and insulating curves, $H_c\approx 99.5$ mT, is marked by the arrow. For fields lower than $H_c$ the sample becomes superconducting whereas for fields higher than $H_c$ the sample becomes insulating.}
		\label{SITT}
\end{figure}

Upon further increase of magnetic field, the tendency to a superconducting state is suppressed, and an insulating behavior appears. As is clearly seen in Fig.~\ref{SITT}, the transition from a superconducting to insulating behavior occurs at a critical resistance of $\sim 6.0$k$\Omega$, which within the error bars of determining the sample's sheet resistance coincides with the quantum of resistance $\rho_Q=h/4e^2\approx 6.5$ k$\Omega$.

\begin{figure}[ht]
	\centering
	\includegraphics[width=1.0\columnwidth]{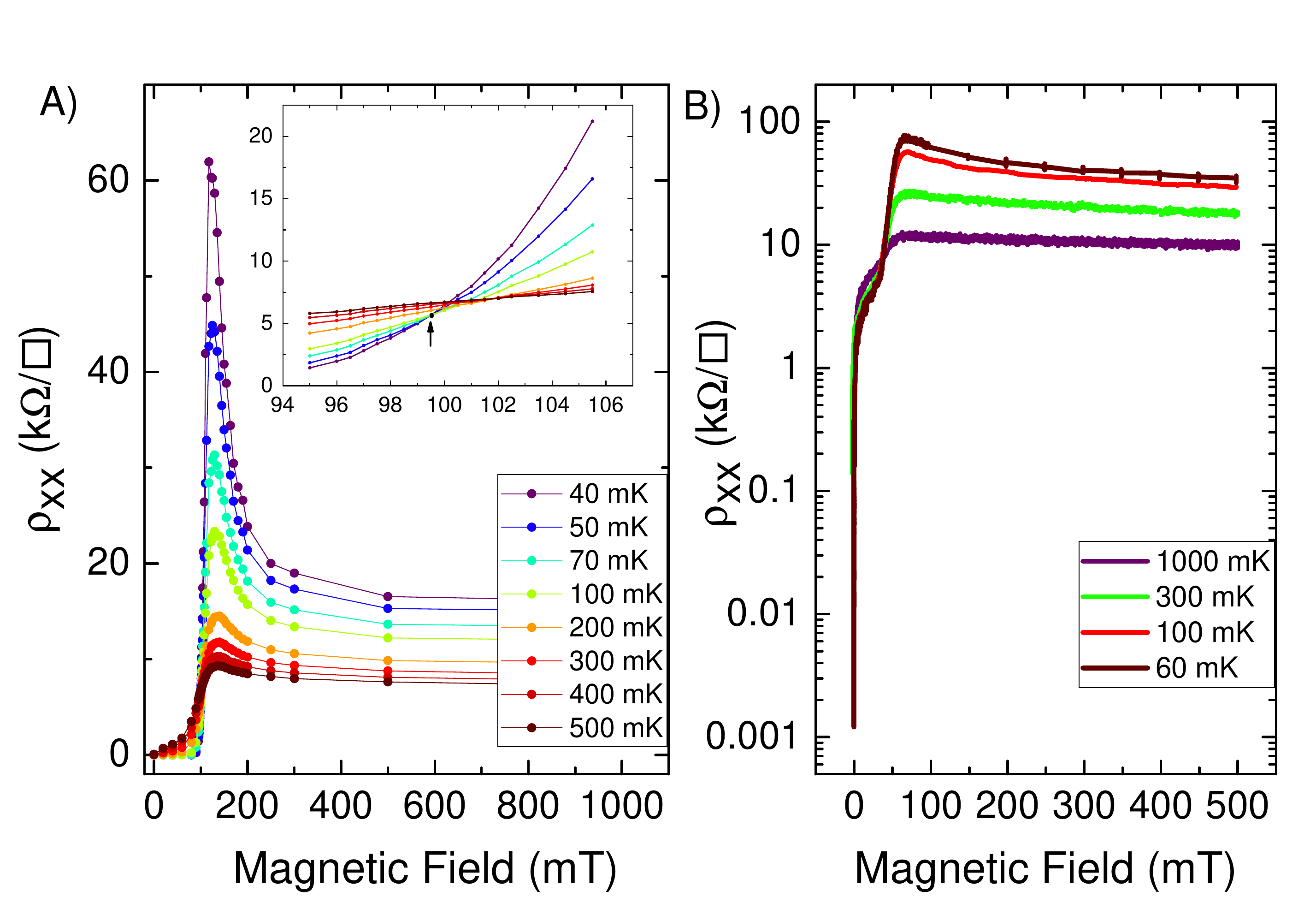}
	\caption{Magnetoresistance at different temperatures of (A) Sample1-\nth{2} anneal and (B) Sample2-\nth{1} anneal. Both of the samples exhibit H-SIT with a well pronounced peak in the insulating phase. The magnetoresistance of Sample1-\nth{2} anneal in (A) was deduced from Fig.~\ref{SITT}.The inset zooms in the critical field region where the curves intersect which is a hallmark of two-dimensional quantum-SIT. $H_c$ for the low temperature range of $\lesssim 100mK$ is marked by black dot and arrow.}
		\label{SITH}
\end{figure}

Identifying the quantum-SIT in temperature dependence data, we can further investigate this transition by studying the magnetoresistance of the samples. A hallmark of such plots is a crossing point that is clearly seen in Fig.~\ref{SITH}. Such a crossing point is a hallmark of two-dimensional quantum-SIT because the upfront amplitude of the resistance scaling function does not depend on any dimension of the sample. The scaling function itself must reflect the interplay between the correlation length that controls the SIT and the thermal length that cuts-off the dynamics of the system.
Further examination of the crossing point in the inset of Fig.~\ref{SITH}(A) shows a progressive drift towards the low temperatures value, where below $\sim 100$ mK they all cross at a critical field that in agreement with the inset of Fig.~\ref{SITT} seems to be bound between $99.25$ mT and $99.75$ mT.\\

\noindent {\it Data Analysis and Discussion} -  Studies of H-SIT in amorphous indium-oxide films have been a subject of intense research efforts for over 15 years, owing primarily to the giant positive magnetoresistance observed on the insulating side of the transition \cite{Gantmakher1998,Sambandamurthy2004,Steiner2005}. Since superconductivity originates from a normal state with a (fermionic) sheet resistance of order a few k$\Omega/\Box$, the observation of a magnetoresistance peak in excess of $\sim$G$\Omega/\Box$ \cite{Sacepe2015} must be interpreted as a consequence of localization of persisting Cooper-pairs \cite{Fisher1990a}. Such a behavior is intrinsic to an array of Josephson-coupled grains, and was demonstrated, e.g. in an ordered array of aluminum islands coupled by controllable system based on a two-dimensional electron gas (2DEG) formed in a semiconductor heterostructure \cite{Bottcher2018}. We may view the H-SIT in the In/InOx system that we report here as a `random version' of the B{\o}ttcher {\it et al.} Al/2DEG system \cite{Bottcher2018}, where here the indium grains have relatively broad grain-size distribution, and are randomly distributed. The underlying InOx that serves as the coupling layer is estimated to have a normal-state sheet resistance of $\sim 10$k$\Omega/\Box$, which could support weak local pairing, particularly if proximitized from the indium grains. Assuming a H-SIT governed by a diverging correlation length $\xi \propto |H-H_c|^{-\nu}$ on either side of the critical field, $H_c$, and invoking a thermal length $L_T\sim (k_BT)^{-1/z}$ associated with the finite temperature that cuts-off the quantum dynamics of the transition ($z$ is the dynamical exponent), it is easy to arrive to a simple expression for the system's sheet resistance \cite{Fisher1990a} near criticality:
\begin{equation}
\rho(H,T)=\rho_c\mathcal{F}\left(|H-H_c|T^{-1/z\nu}\right)
\label{scale}
\end{equation}
where $\mathcal{F}(x)$ is a scaling function that depends on the distance to the critical field. Requiring that $\rho(H_c,0)=\rho_c$, we must have $\mathcal{F}(0)=1$, independent of temperature, which is a manifestation of the emerging crossing point in the measured field-dependent resistance isotherms.

\begin{figure}[ht]
	\centering
	\includegraphics[width=1.0\columnwidth]{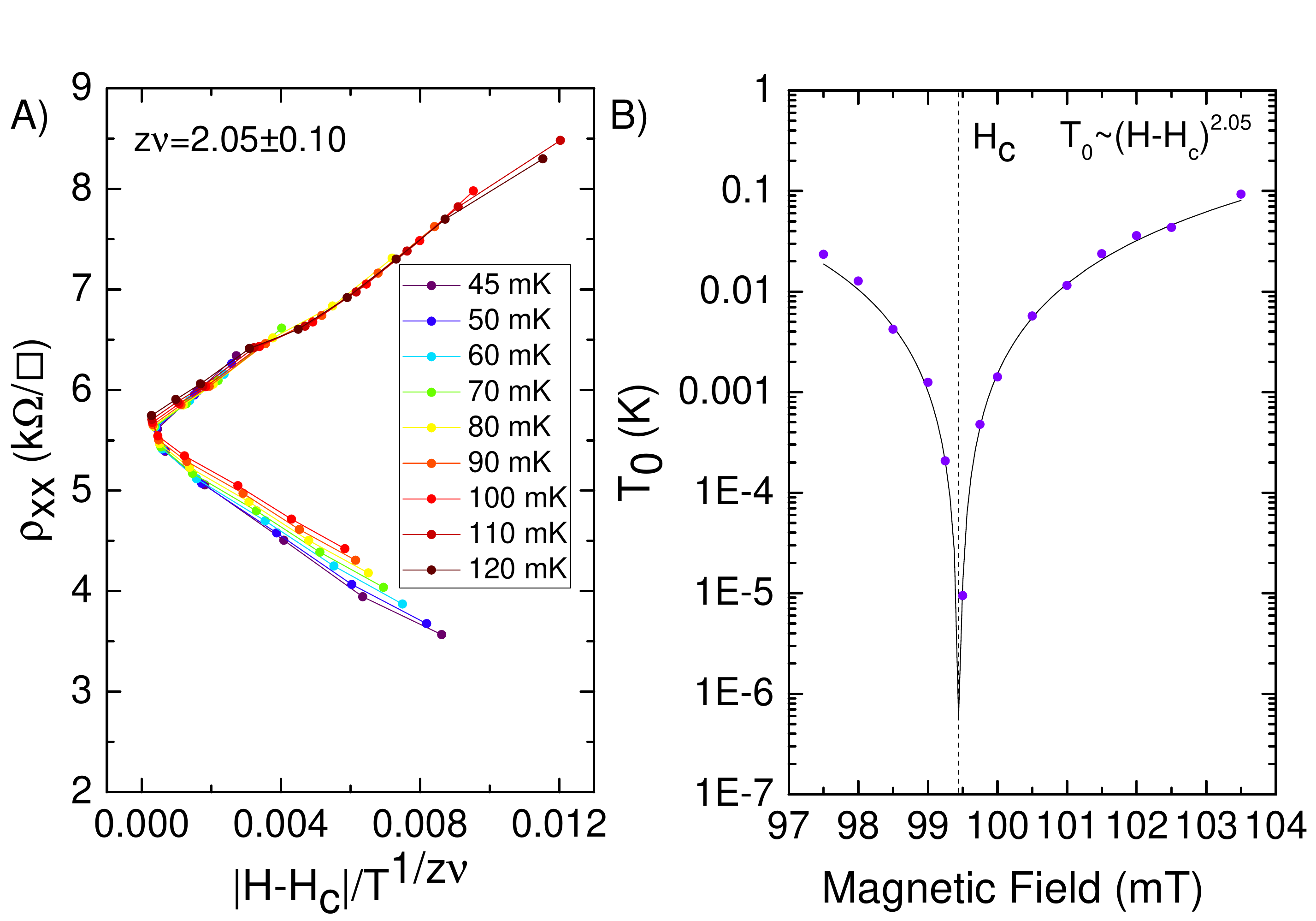}
	\caption{Verification of the scaling laws of the SIT in Sample1-\nth{2} anneal. a) Scaling of isotherms near $H_c$ for $z\nu=2.05$. b) The characteristic temperatures extracted from the fits to Eqs.~\ref{pairs} and~\ref{vortex} as a function of magnetic field, showing a critical behavior at $H_c$
consistent with the determined critical exponent.}
		\label{Scaling}
\end{figure}

For further analysis we focus on the data obtained for Sample1-\nth{2} anneal which is the closest to the SIT from the superconductor side exhibiting zero resistance at zero magnetic field.
To apply the above scaling in a consistent way we go back to  Fig.~\ref{SITT}, to determine that the isotherms near $H_c\approx 99.5$ mT, seem to reach their low-temperature asymptotic value below $\sim 100$ mK. Fig.~\ref{Scaling}(A) is a plot of the low temperature data fitted to the form of Eqn.~\ref{scale}. The data collapse was achieved by adjusting the critical field and the complex exponent $z\nu$. Within the choice of temperature range and reduced magnetic field, we find that the data is best fit with $H_c=99.4 \pm 0.1$ mT and $z\nu=2.05 \pm 0.10$. We note however that if we restrict ourselves to a much smaller reduced parameter, $x\lesssim 0.005$, the quality of the data is not good enough to rule out exponents as large as $z\nu\sim 2.5$.
Thus to further investigate the critical behavior we use a different test of the scaling near the H-SIT, which was recently introduced by Breznay {\it et al.} \cite{Breznay2016}.
Starting on the insulating side of the transition, we consider it most likely that the resistance is dominated by variable-range-hopping (VRH) of Cooper pairs.  In order to have a simple expression which connects smoothly with the scaling analysis at the H-SIT (Eqn.~\ref{scale}), we take
\begin{equation}
\rho_{xx}(T,H) \approx \rho_c\exp[(T_0(H)/T)^{\delta}] \ {\rm for} \ H>H_c
\label{pairs}
\end{equation}
with $\delta$ describing the nature of the VRH that govern the electronic transport.

On the superconducting side, for $H<H_c$, we expect that the longitudinal resistivity will be dominated by motion of vortices. Assuming Cooper pairs-vortex duality, the (measured) conductivity tensor $\underline{\sigma}$ is related to the vortex-resistivity tensor $\underline{\rho}^v$ according to $\underline{\sigma}=(4e^2/h)^2 \underline{\rho}^v$, implying that  \cite{Auerbach2006}:
\begin{equation}
\rho_{xx}=(4e^2/h)^2\sigma_{xx}^v =\rho_c\exp[-(T_0(H)/T)^{\delta}],
\label{vortex}
\end{equation}
where we expect vortex VRH below $H_c$ to replace Cooper-pairs VRH above $H_c$ \cite{Breznay2016}. Consistency between Eqns.~\ref{scale}, \ref{pairs}, and \ref{vortex} requires that $T_0(H)\propto |H-H_c|^{z\nu}$. Fig.~\ref{Scaling}(B) is a result of fitting Eqns.~\ref{scale} and \ref{pairs} to the data and extracting the characteristic temperature-scale $T_0(H)$. The fit yields the same $H_c=99.4$ mT and a scaling exponent $z\nu=2.05\pm0.10$ as in Fig.~\ref{Scaling}(A).

The study of $\rho_{xx}$ on the insulating side also allows for the detection of the transition to the non-equilibrium state where electrons fail to fully thermalize with the lattice. To observe that transition we continue to assume the form of Eqn.~\ref{pairs} and thus plot ${\rm ln}[{\rm ln}(\rho_{xx}/\rho_c)]$ vs. ${\rm ln}(1/T)$. Such a plot allows to confirm the exponent $\delta\approx 1/2$ discussed above (which was determined in a direct fit to Eqn.~\ref{pairs}), and it also allows to observe the increase of that exponent at low temperatures, away from the H-SIT.
\begin{figure}[ht]
	\centering
	\includegraphics[width=1.0\columnwidth]{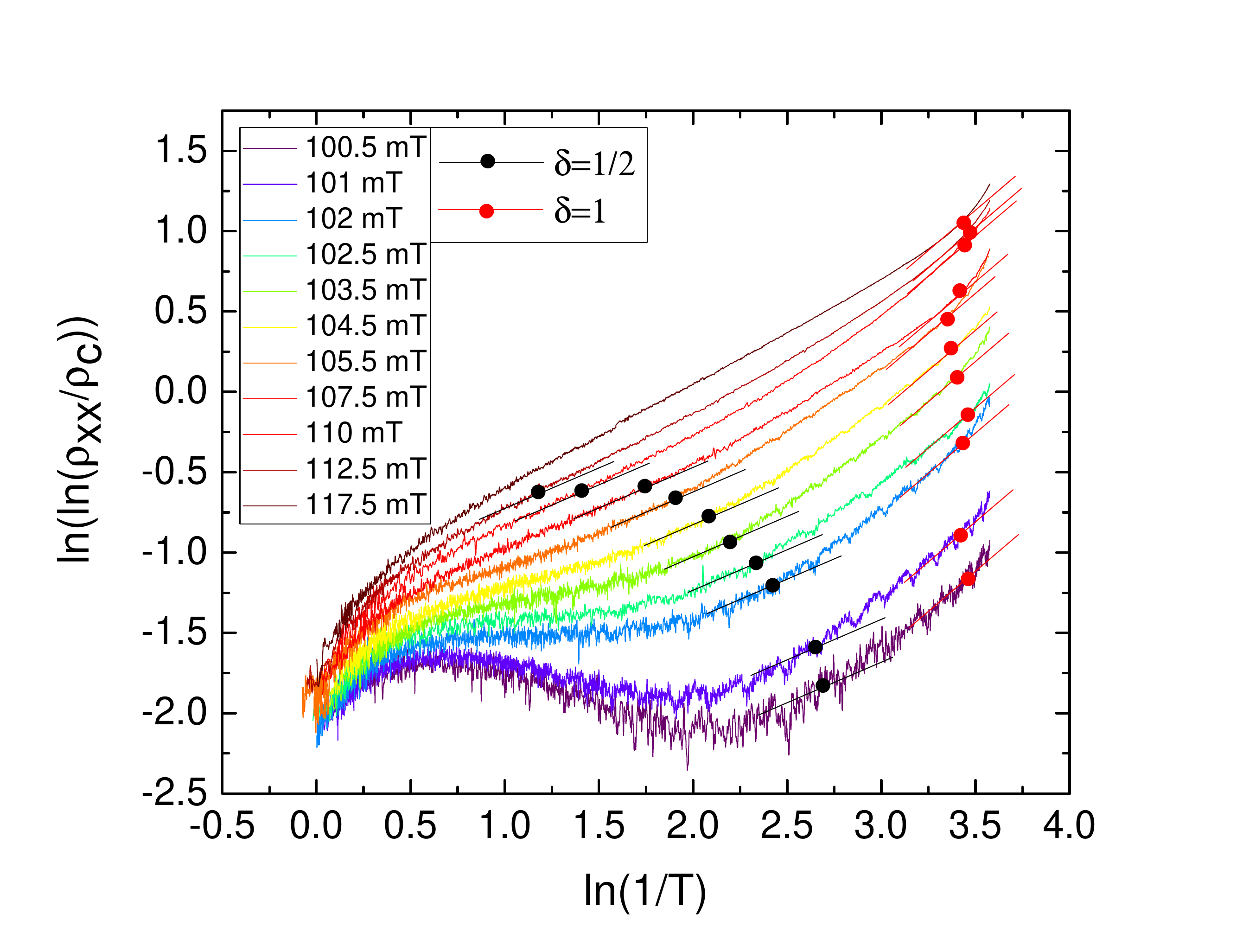}
	\caption{A rough determination of the crossover temperature from equilibrium VRH behavior to nonequilibrium state of Sample1-\nth{2} anneal. Black circles denote the first deviation from VRH with $\delta =1/2$, whereas red circles denote the crossover to ``super-activation'' with $\delta > 1$.}
		\label{Nonequilibrium}
\end{figure}
Fig.~\ref{Nonequilibrium} depicts this evolution for magnetic fields above $H_c$. As the temperature is reduced, a gradual crossover with an intermediate ``simple activation'' behavior with $\delta=1$ is observed, followed by a ``super-activation''  behavior where a strong decoupling of the electronic system from the lattice occurs with $\delta > 1$. While the deviation from VRH depends on the proximity to the H-SIT.  It is important to note that the initial deviation from VRH depends on temperature - the closer the field is to $H_c$ (and hence the resistance to $\rho_c$), the lower the temperature of that crossover. However, the full transition to ``super-activation''  is no longer dependent on the proximity to the H-SIT, but is rather an intrinsic property of the sample, where for the sample shown in Fig.~\ref{Nonequilibrium} it occurs at $T_{Neq}\sim 30$ mK.

\noindent {\it Summary and Conclusions} - In this paper we introduced a novel material system for the study of magnetic-field tuned superconductor to insulator transition. Utilizing an underlying amorphous indium-oxide layer to control the coupling between a subsequently deposited thin films of pure indium islands, we create a unique granular system where the coupling between the grains can be controlled by careful annealing of the composite system in vacuum. Tuning the coupling to lie just at the borderline where superconductivity in the underlying InOx is suppressed, which is also close to the metal-insulator transition of the InOx, the hybrid system exhibits a ``giant'' magnetoresistance above the H-SIT, with critical behavior that manifests the duality between Cooper pairs and vortices. We further study the transition to a non-equilibrium state on the insulating side of the H-SIT, observing its shift towards $H_c$  at lower temperatures.
\\

\noindent {\bf Acknowledgments:} We acknowledge important discussions with Boris Spivak, and Steven Kivelson. Work at Stanford University was supported by the National Science Foundation Grant NSF-DMR-1808385. Work at Tel-Aviv University was supported by the Israeli Science Foundation (Grant No. 277/16), the Israel Ministry of Science and Technology (Contract No. 67091), the US-Israel Binational Science Foundation (Grant No. 2014098), and the Wolfson Applied Materials Research Centre (Tel-Aviv University).

\end{document}